\documentclass[12pt]{article}
\usepackage{epsf}
\def\beq{\begin{equation}}
\def\eeq{\end{equation}}
\def\ap#1#2#3 {Ann. Phys. (NY) {\bf#1} (19#2) #3}
\def\err#1#2#3 {{\it Erratum} {\bf#1} (19#2) #3}
\def\ib#1#2#3 {{\it ibid.} {\bf#1} (19#2) #3}
\def\ijmp#1#2#3 {Int. J. Mod. Phys. {\bf#1} (19#2) #3}
\def\jetp#1#2#3 {JETP Lett. {\bf#1} (19#2) #3}
\def\mpl#1#2#3 {Mod. Phys. Lett. {\bf#1} (19#2) #3}
\def\np#1#2#3 {Nucl. Phys. {\bf#1} (19#2) #3}
\def\pl#1#2#3 {Phys. Lett. {\bf#1} (19#2) #3}
\def\prep#1#2#3 {Phys. Rep. {\bf#1} (19#2) #3}
\def\prev#1#2#3 {Phys. Rev. {\bf#1} (19#2) #3}
\def\prl#1#2#3 {Phys. Rev. Lett. {\bf#1} (19#2) #3}
\def\sjnp#1#2#3 {Sov. J. Nucl. Phys. {\bf#1} (19#2) #3}
\def\spj#1#2#3 {Sov. Phys. JETP {\bf#1} (19#2) #3}
\def\spu#1#2#3 {Sov. Phys. Usp. {\bf#1} (19#2) #3}
\def\zp#1#2#3 {Zeit. Phys. {\bf#1} (19#2) #3}

\begin{document}
\begin{titlepage}
\begin{center}
{\Large \bf Theoretical Physics Institute \\
University of Minnesota \\}  \end{center}
\vspace{0.2in}
\begin{flushright}
TPI-MINN-03/11-T \\
UMN-TH-2136-03 \\
April 2003 \\
\end{flushright}
\vspace{0.3in}
\begin{center}
{\Large \bf  Remarks on the decays $\chi_{bJ}^\prime \to \omega \,
\Upsilon$
\\}
\vspace{0.2in}
{\bf M.B. Voloshin  \\ }
Theoretical Physics Institute, University of Minnesota, Minneapolis,
MN
55455 \\ and \\
Institute of Theoretical and Experimental Physics, Moscow, 117259
\\[0.2in]
\end{center}

\begin{abstract}
The recently observed by CLEO cascade decays $\Upsilon(3S) \to \gamma \,
\chi_{bJ}^\prime \to \gamma \, \omega \, \Upsilon$ are discussed. It is
shown that within the nonrelativistic description of bottomonium in the
heavy $b$ quark limit, the ratio of the rates of the transitions
$\chi_{bJ}^\prime \to \omega \, \Upsilon$ with $J=1$ and $J=2$ should be
given by the ratio of the corresponding $S$ wave phase space factors. As
a result, the rate of the observed cascade transitions through the
$\chi_{b2}^\prime$ resonance should be close to that through the
$\chi_{b1}^\prime$. It is suggested that the ratio of the discussed
cascade rates can also be tested by measuring a simple angular
correlation.
\end{abstract}

\end{titlepage}

The CLEO collaboration has recently presented data\cite{cleo} indicating
an observation of the decay cascade $\Upsilon(3S) \to \gamma \,
\chi_{bJ}^\prime \to \gamma \, \omega \, \Upsilon$, involving the
hadronic transitions $\chi_{bJ}^\prime \to \omega \, \Upsilon$. The
kinematical analysis of the data favors dominance of transitions trough
the $\chi_{b1}^\prime$ resonance with the branching ratio
$B(\chi_{b1}^\prime \to \omega \, \Upsilon) = 2.3 \pm 0.4 \%$. It is the
purpose of this letter to point out that within the standard description
of the hadronic transitions between the levels of the $b \bar b$ system
it would be rather troublesome to accommodate a dominance of the $J=1$
$\chi_{b1}^\prime$ resonance over that with $J=2$, $\chi_{b2}^\prime$.
Namely, the expected ratio of the absolute decay rates,
$\Gamma(\chi_{b2}^\prime \to \omega \, \Upsilon)/
\Gamma(\chi_{b1}^\prime \to \omega \, \Upsilon)$, is given by the ratio
of the $S$ wave phase space factors, which is approximately 1.4. Thus
allowing for the relative rate of the initial radiative transitions from
the $\Upsilon(3S)$ resonance and also for the ratio of the total decay
rates of $\chi_{b2}^\prime$ and $\chi_{b1}^\prime$, one would estimate
the relative contribution of these resonances to the observed rate of
the cascade decays approximately as
\beq
r_{2/1} \equiv { \Gamma(\Upsilon(3S) \to \gamma \, \chi_{b2}^\prime \to
\gamma \, \omega \, \Upsilon) \over \Gamma(\Upsilon(3S) \to \gamma \,
\chi_{b1}^\prime \to \gamma \, \omega \, \Upsilon)} \approx 1.3 \pm
0.3~,
\label{r21}
\eeq
where the indicated uncertainty arises mainly from the present knowledge
of the ratio of the total decay rates of the $\chi_{bJ}^\prime$
resonances.

Furthermore, in the nonrelativistic limit of heavy $b$ quarks, where the
spin of the heavy quark effectively decouples, the spin structure of the
decay amplitudes $\chi_{b1}^\prime \to \omega \, \Upsilon$ and
$\chi_{b2}^\prime \to \omega \, \Upsilon$ is fully fixed, which leads to
a distinct prediction for the angular correlation, different for each of
these decays, between the observed decay products of $\omega$ and
$\Upsilon$. Thus an additional information on the ratio $r_{2/1}$ can be
found in the already existing data by studying the angular correlation
between the direction determined by the leptons ($e$ or $\mu$) emerging
from $\Upsilon \to \ell^+ \, \ell^-$ and the plane determined by the
momenta of the charged pions from the decay $\omega \to \pi^+ \, \pi^-
\, \pi^0$.

In order to analyze the amplitudes of the decays $\chi_{bJ}^\prime \to
\omega \, \Upsilon$ one can recall that the hadronic transitions of the
discussed type are viewed as proceeding through emission of a soft
gluonic field in the heavy quarkonium transition, which field
materializes as light hadron(s)\cite{gottfried}, the $\omega$ resonance
in the case considered here. For nonrelativistic heavy $b$ quarks the
dominant interaction with a soft gluon field is that with the
chromoelectric component ${\vec E}^a$, since the interaction with the
cromomagnetic field is suppressed by an inverse power of the heavy mass,
$m_b^{-1}$. It is important for the present discussion that the quantum
numbers of the $\omega$ resonance ($J^{PC}=1^{--}$) make possible the
$\omega$ production by exclusively the chromoelectric gluon field i.e.
without the need for its chromomagnetic component. The
amplitude of such production can be described by a generic symmetric
amplitude (in the rest frame of $\omega$)
\beq
\langle \omega ({\vec \epsilon}) |d_{abc}\, \int \, f(q_1, q_2, q_3)\,
\left ({\vec E}^a(q_1) \cdot {\vec E}^b(q_2)\right ) \, {\vec E}^c(q_3)
\, \delta^{4} (q_1+ q_2+ q_3 - p) \, dq_1 \, dq_2 \, dq_3 |0 \rangle =
A_\omega \, {\vec \epsilon}\,^*~,
\label{aom}
\eeq
where ${\vec \epsilon}$ stands for the polarization amplitude of
$\omega$, $p=(m_\omega, {\vec 0})$ is its four-momentum, $d_{abc}$ are
the symmetric SU(3) constants, and the form factor $f(q_1, q_2, q_3)$ is
totally symmetric in its arguments and is determined by (presently
poorly known) details of the dynamics of the transition between the
heavy quarkonium levels.

In the limit where the heavy quarkonium is considered as a compact
object and only its chromoelectric dipole interaction with ${\vec E}^a$
is retained, the form factor $f$ would be reduced to a point-like in the
coordinate space. However it is known that such approximation does not
work well for hadronic transitions from the $\Upsilon(3S)$ resonance, in
particular for the decay $\Upsilon(3S) \to \pi \, \pi \, \Upsilon$ (for
a discussion see e.g. the review \cite{vz}, the latest data are
presented in the talk\cite{cleo}), which behavior might be attributed
to a rather high excitation energy of the $\Upsilon(3S)$. Since the
$\chi_{bJ}^\prime$ resonances are only slightly below the
$\Upsilon(3S)$, an application of the leading terms in the multipole
expansion to hadronic transitions from them can be somewhat questionable
as well. Fortunately, the assumption of the applicability of this limit
is not required for the considerations presented here. What is really
important is that it is sufficient to consider the interaction of the
heavy quarks only with the chromoelectric component of the gluon field.
In the leading nonrelativistic order this interaction depends only on
the coordinate variables of the heavy quarks and does not depend on
their spin. In the same order the wave functions of the heavy quarkonium
states can be factorized into the spatial part, $\psi$, and the spinor
part $\zeta$. For the $\Upsilon$ resonance ($^3S_1$ state) the spatial
part is a scalar $\psi_0$ and the spinor part is a vector, ${\vec
\zeta}_0$, while for the $\chi_{bJ}^\prime$ resonances ($^3P_J$ states)
the spatial part transforms as a vector ${\vec \psi}_1$ and the spinor
part is a vector too: ${\vec \zeta}_1$. Since, as discussed, the spin of
the $b$ quarks decouples from the emission of $\omega$, the $S$
wave\footnote{It can be noticed that the fact that the discussed
transitions are due to the dominant chromoelectric interaction and can
proceed in the $S$ wave somewhat makes up for the `Zweig rule'
suppression, in a semi-quantitative agreement with the
noticeable observed rate.}  amplitude of the transitions from the
$\chi_{bJ}^\prime$ states can be written as
\beq
A(\chi_{bJ}^\prime \to \omega \, \Upsilon)= A\, \epsilon^*_i
\psi_0^* \, \zeta_{0j}^* \, P^{(J)} \, \psi_{1i} \, \zeta_{1j}~,
\label{ampl}
\eeq
where $P^{(J)}$ is the projector of the direct product $\psi_{1i} \,
\zeta_{1j}$ on the states with definite total spin $J$.
Clearly, the higher partial wave amplitudes for the emission of $\omega$
can be neglected, given that the transitions from the $\chi_{b1}^\prime$
and $\chi_{b2}^\prime$ are just above the threshold\cite{pdg}: $\Delta_1
\equiv M(\chi_{b1}^\prime)-M(\Upsilon)-m_\omega = 12.3 \pm 0.6 \, MeV$,
$\Delta_2 \equiv M(\chi_{b2}^\prime)-M(\Upsilon)-m_\omega = 25.6 \pm 0.5
\, MeV$ (while the $\chi_{b0}^\prime$ resonance is well below the
kinematical threshold for the discussed transition).

The fact that the dependence on the spin of the $b \bar b$ pair in
eq.(\ref{ampl}) factorizes out, implies that the ratio of the transition
rates from the $J=2$ and $J=1$ states is given by the ratio of the $S$
wave phase space factors:
\beq
R_{2/1} \equiv {\Gamma(\chi_{b2}^\prime \to  \omega \, \Upsilon) \over
\Gamma(\chi_{b1}^\prime \to  \omega \, \Upsilon)}= \sqrt{{\Delta_2 \over
\Delta_1}} \approx 1.4~.
\label{rgam}
\eeq

The ratio $r_{2/1}$ of the rates for the cascade processes can then be
found as
\beq
r_{2/1}={\Gamma(\Upsilon(3S) \to \gamma \, \chi_{b2}^\prime) \over
\Gamma(\Upsilon(3S) \to \gamma \, \chi_{b1}^\prime)} \,
{\Gamma_{tot}(\chi_{b1}) \over \Gamma_{tot}(\chi_{b2})} \, R_{2/1}~.
\label{r21t}
\eeq
The ratio of the radiative decay rates from the $\Upsilon(3S)$ can be
determined from the standard dipole-emission formula
$\Gamma(\Upsilon(3S) \to \gamma \, \chi_{bJ}^\prime) \sim (2J+1) \,
k_\gamma^3$, which results in $\Gamma(\Upsilon(3S) \to \gamma \,
\chi_{b2}^\prime) / \Gamma(\Upsilon(3S) \to \gamma \, \chi_{b1}^\prime)
\approx 1.1\,$. The experimental value for this ratio\cite{pdg} is
consistent with the theoretical number, but has a considerably larger
uncertainty. A theoretical estimate of the ratio of the total decay
rates of the $\chi_{b2}^\prime$ and $\chi_{b1}^\prime$ is plagued by the
well known infrared sensitivity of the annihilation widths\cite{bgr,
nosvvz}. This ratio however can be evaluated as an inverse of the
relative branching ratia $B(\chi_{bJ} \to \gamma \,
\Upsilon)\,$\footnote{The data on the ratio of $B(\chi_{bJ} \to \gamma
\, \Upsilon(2S))$ could be used as well. However the current
experimental uncertainty in these data is larger than for the radiative
transitions to $\Upsilon(1S)$.} after a minor correction due to a slight
difference in the factors $k_\gamma^3$: $\Gamma_{tot}(\chi_{b1}) /
\Gamma_{tot}(\chi_{b2}) \approx 0.8 \pm 0.15\,$. Thus collecting the
factors in eq.(\ref{r21t}), one arrives at the estimate of the cascade
rate ratio $r_{2/1}$ given in eq.(\ref{r21}).

Besides a slightly different kinematics in the discussed cascade
transitions through the $J=2$ and $J=1$ resonances, their contribution
can be additionally resolved by measuring the angular correlation
between the decay products of the $\Upsilon$ and $\omega$ resonances.
Indeed, these resonances are observed\cite{cleo} through their decays
$\Upsilon \to \ell^+ \, \ell^-$ and $\omega \to \pi^+ \, \pi^- \,
\pi^0$. In the rest frame of the $\Upsilon$ resonance the two leptons
are emitted `back to back' thus defining one direction, and the other
direction is the perpendicular to the plane defined by the charged pion
momenta $p_+$ and $p_-$. Defining the angle $\theta$ as that between
these two directions, one can find the distribution of the rates over
$\cos \theta$ for the cascade transitions through the $\chi_{b2}^\prime$
and through the $\chi_{b1}^\prime$, neglecting the motion of the
$\omega$ in the rest frame of $\Upsilon$. (The small corrections due to
this motion are proportional to $\Delta_{1,2}/m_\omega$ and are
parametrically of the same order as that due to the contribution of
higher partial wave(s) in the amplitudes of $\chi_{bJ}^\prime \to \omega
\, \Upsilon$.) Since the structure of the decay amplitudes for
$\Upsilon$ and $\omega$ is uniquely determined: $A(\Upsilon \to \ell^+
\, \ell^-) \propto {\vec \zeta_0} \cdot ({\overline \ell} {\vec \gamma}
\ell)$ and $A(\omega \to \pi^+ \, \pi^- \, \pi^0) \propto {\vec
\epsilon} \cdot ({\vec p}_+ \times {\vec p}_-)$, the corresponding
angular distributions are readily found from eq.(\ref{ampl}) as
\beq
{d \Gamma(\chi_{b2}^\prime) \over d \cos \theta}  \propto \left ( 1 -{1
\over 7} \, \cos^2 \theta \right )~,~~~~~~~ {d \Gamma(\chi_{b1}^\prime)
\over d \cos \theta}  \propto \left ( 1 +  \cos^2 \theta \right )~.
\label{ad}
\eeq
The correlation coefficient $a$ in the overall angular distribution in
the cascade $\Gamma(\Upsilon(3S) \to \gamma \, \chi_{bJ}^\prime \to
\gamma \, \omega \, \Upsilon) \propto (1+a \, \cos^2 \theta)$ with the
two branches, corresponding to $J=2$ and $J=1$, mixed in the proportion
determined by $r_{2/1}$, is therefore found as
\beq
a={5-r_{2/1} \over 5+ 7 \, r_{2/1}}~.
\label{corr}
\eeq
The expected value of the ratio $r_{2/1}$ in eq.(\ref{r21}) corresponds
to $a \approx 0.26 \pm 0.06$, while in the limit of a strong dominance
of the cascade through $\chi_{b1}^\prime$, i.e. when $r_{2/1} \to 0$,
the correlation coefficient obviously equals one. Hopefully, the already
available data are sufficient for resolving between these two values.

I thank Jon Urheim, who brought to my attention the data\cite{cleo}, for
enlightening discussion of the experimental situation and I gratefully
acknowledge a useful discussion of the theoretical aspects with Arkady
Vainshtein. This work is supported in part by the DOE grant
DE-FG02-94ER40823.

\end{document}